\documentclass{article}
\usepackage{graphicx} 

\title{Treatment, evidence, imitation, and chat
}
\author{samweisenthal}
\date{July 2025}

\usepackage{amssymb}
\usepackage{amsmath}
\usepackage{amsthm}
\usepackage{xcolor}
\usepackage{hyperref}
\usepackage[numbers]{natbib}
\usepackage[a4paper, total={6in, 10in}]{geometry}
\usepackage{authblk}
\usepackage{setspace}
\usepackage[english]{babel}
\newtheorem{theorem}{Heuristic}

\author{
Samuel J. Weisenthal\\  \href{mailto:samweisenthal@gmail.com}{samweisenthal@gmail.com}
}

\begin{document}

\maketitle

\affil[1]{University of Rochester Medical Center, 601 Elmwood Ave, Rochester, NY, 14642
}
\affil[1]{samuel\_weisenthal@urmc.rochester.edu}

\section*{Abstract}
Large language models are thought to have the potential to aid in medical decision making. This work investigates the degree to which this might be the case.  We start with the treatment problem, the patient's core medical decision-making task, which is solved in collaboration with a clinician. We discuss different approaches to solving it, including, within evidence-based medicine, experimental and observational data. We then discuss the chat problem, and how this differs from the treatment problem---in particular with respect to imitation (and how imitation alone cannot solve the true treatment problem, although this does not mean it is not useful). We then discuss how a large-language-model-based system might be trained to solve the treatment problem, highlighting that the major challenges relate to the ethics of experimentation and the assumptions associated with observation. We finally discuss how these challenges relate to evidence-based medicine and how this might inform the efforts of the medical research community to solve the treatment problem. Throughout, we illustrate our arguments with the cholesterol medications, statins.

\section{Introduction}
In recent years, we have seen unprecedented gains in chatbot capabilities, owing largely to reinforcement learning-guided large language models (LLMs) \citep{radford2018improving, brown2020language}, such as ChatGPT \citep{chatGPT}.  Although such technologies are already used in healthcare (e.g., as digital scribes \citep{van2024impact,haberle2024impact}, there is hope that they might one day also provide clinical decision support or automation.  This hope might be related to the performance of large language models on medical licensing exams \citep{kumah2023chatgpt, fijavcko2023can, kung2023performance, ayers2023evaluating, goodman2023accuracy} or, despite some unfavorable results   \citep{thirunavukarasu2023large, thirunavukarasu2023trialling, sarraju2023appropriateness, weisenthal2023chatgpt, narayanan2024ai,hager2024evaluation, bean2026reliability, ramaswamy2026chatgpt}, studies on language models and clinical reasoning in other scenarios \citep{johri2025evaluation, brodeur2024superhuman, wu2025large,
10.1001/jama.2023.8288, 
10.1093/jamia/ocad072, ah2025large, 
gumilar2024assessment, rodman2023artificial, kweon2024ehrnoteqa, li2024mediq,
wang2024direct, wu2024medjourney}.
  
In this work, we discuss the potential for chatbots to aid with treatment decisions.  We begin by describing the main concern of the patient, which we call the treatment problem. We show how it can be represented within an expected-utility, decision-analytic \citep{pauker1987decision} framework, which is closely related to reinforcement learning \citep{sutton2018reinforcement, almudevar2014approximate}. 
We describe some existing approaches to address the treatment problem, including, as related to trials and observational studies, evidence-based medicine itself.  Expanding on the latter, we also discuss dynamic treatment regimes (see, e.g., \citep{murphy2003optimal, 
luckett2019estimating, Chakraborty2013, yazzourh2025medical, laber2014dynamic}), which are closely related to off-policy reinforcement learning (see, e.g., \citep{futoma2020popcorn, raghu2017deep, raghu2017continuous, gottesman2020interpretable, 
ertefaie2018constructing, 
sutton2018reinforcement, 
ghasemi2025personalized,
jonsson2019deep, petersen2019deep,gottesman2018evaluating, yu2021reinforcement}).   We  then discuss imitation learning \citep{torabi2018behavioral, kober2010imitation} and how it can be combined with utility maximization \citep{peters2010relative, levine2014learning, schulman2015trust, relsparSIM, achiam2017constrained, kallus2020efficient}.

We then discuss the chat problem, and how it balances imitation and utility optimization \citep{ouyang2022training}, and the implications of this tradeoff in terms of solving the treatment problem. We caution against imitation of medical notes.  We finally describe how a language model might be trained to solve the treatment problem directly, arguing that such a task requires one to overcome the core challenges in evidence-based medicine. We comment on the implications of this inter-relatedness between chatbots and evidence-based medicine, and how it might inform the medical research community's efforts to solve the treatment problem more generally.

\section{The treatment problem}

The treatment problem can be addressed using utility optimization, evidence-based medicine, experience, knowledge, or imitation, as will be described below.

\subsection{Patient utility optimization}

The treatment problem can be formulated as a decision analysis \citep{pauker1987decision, lindley1991making} or reinforcement learning \citep{sutton2018reinforcement} task. In particular, we have, for patient characteristics, $X,$  treatment, $T,$ and ``policy", $\pi^*,$ that
\begin{align}
\label{eq:treatmentpolicy}
T \sim {\pi^*}(T|X),
\end{align}
where
\begin{align}
\label{eq:treatmentopt}
\pi^* = \arg\max_{\pi} \mathbb{E}_{\pi}U (T, X),
\end{align}
where $U$ is the patient's utility, and where $\mathbb{E}_{\pi}U$ stands for the patient’s expected utility, if the policy $\pi$ were used.  

In more detail, 
\begin{align}
\label{eq:detail}
\mathbb{E}_{\pi}U&=\int udP_{\pi}(u)\\&=\int\int\int  u dP(u|do(t),x)d\Pi(t|x)dP(x)
\end{align} 
with $\Pi$ the CDF of $\pi$ and  $\int zdP(z)$ defined, for CDF $P,$ as $\int z p(z)dz,$ if $z$ is continuous, and $\sum_z zp(z),$ if $z$ is discrete. 
 Note that the expectation in (\ref{eq:treatmentopt}) is interventional \citep{pearl2009causality}, and hence the $do$ notation within $dP(u|do(t),x),$ which can be conceptualized, for example, as the estimand of a randomized controlled trial (note that this kind of causal reasoning goes beyond correlation, and this has substantial implications for the development of intelligent computer systems \citep{marcus2018deep, bareinboim2024introduction}). Note that $X$ does not necessarily correspond to the set of confounders. The optimization in  (\ref{eq:treatmentopt}) gives a policy, $\pi^*,$ that solves the treatment problem.

As an example of (\ref{eq:treatmentopt}), suppose we wish to determine whether to recommend a statin to a patient who has certain characteristics, $X,$ where 
\begin{equation}
\label{eq:X.statins}
X=(\text{age, blood pressure, cholesterol levels, smoking status, etc...})^T.
\end{equation}
We hope to maximize the patient's overall utility, $U,$ which will be a function of whether or not the statin is prescribed, $T,$ their post-treatment state, which we can define as $X',$ and will include cardiovascular outcomes (strokes, myocardial infarctions, etc.) and side effects (myopathy, diarrhea, etc.), and, lastly, their resulting utility, $U(X').$

The policy, $\pi,$ in (\ref{eq:treatmentopt}) therefore maps the patient characteristics, $X,$ to a treatment decision, $T$ (which will be, e.g., 1 if the patient is prescribed a statin and 0 otherwise). This policy  will maximize $U(X'),$ in expectation.  Note that sometimes one writes (\ref{eq:detail}) to depend on $X',$ but we suppress  $X'$ here, since $U$ is a function of $X'.$

The utility, $U,$ often varies by patient, which makes solving (\ref{eq:treatmentopt})  quite difficult (i.e., different patients will experience cardiovascular events and side effects differently).  In general, truly solving the treatment problem is essentially intractable for most cases.\footnote{With regard to language models, this may present issues in and of itself \citep{kirichenko2025abstentionbench}}  We can attempt to solve it, however, using tools from evidence-based medicine.  

\subsection{Evidence-based medicine}
\label{sec:TreatevidenceExperience}
Evidence-based medicine allows one to use data to address the treatment problem (\ref{eq:treatmentopt}). The evidence can be derived from trials or observational studies, both of which will be discussed below.

\subsubsection{Trials}
\label{sect:TxTrials}
A trial involves experimentation, i.e. randomization \citep{fisher1970statistical} (for examples of trials, see, e.g., \citet{stone20142013} or \citet{chaudhuri2023use}). For treatment $T,$ patient information $X,$ and utility $U,$ a trial would ideally\footnote{Other approaches, such as using ordinal models, as in \citet{peterson1990partial}, for ranking outcomes by utility, are also possible, but the joint distribution  $P(u|do(t),x)$ is an ideal estimand due to its presence in the expected utility equation.} give an estimate of $P(u|do(t),x).$ This gives the interventional probability of, for example, the patient being well given no treatment, the patient being well given treatment, etc (for statins, this would include the probabilities of cardiovascular events and side effects (such as myopathy) under different treatment choices).  

Given these quantities from a trial, one would be able to plug an estimate of $P(u|do(t),x)$ into (\ref{eq:treatmentopt}) and solve for $\pi.$ For example, in the case of statins, $\pi$ would give a mapping from the patient characteristics---$X,$ which, as discussed in (\ref{eq:X.statins}), contain variables like age and cholesterol---to the treatment choice with respect to statins, $T.$  This policy, $\pi,$ is a direct solution to (\ref{eq:treatmentopt}). 

\subsubsection{Heuristics}
However, a trial does not usually estimate $P(u|do(t),x),$ possibly because it is difficult---from a perspective of statistical power---to estimate something so complex as the joint distribution of utility. The following indirect, heuristic approaches are usually taken instead. 

\begin{theorem}
\label{heur1}
First, one consults a trial for an estimate of a treatment effect, \[\tau=E[Y|do(T=1)]-E[Y|do(T=0)],\] which might correspond to whether a medication would improve some outcome, $Y$ (the outcome, $Y,$ is a component of the patient's post-treatment state, $X',$ i.e., $X'=(Y,\cdots)^T,$ and $Y$ is often only one argument of $U$ (in other words, $U$ is a function of $Y,$ but $U$ also depends on other things in $X',$ such as side effects)). For the statin example, $\tau$ would be an estimate of the treatment effect for statins, which might be derived from a trial in which patients are randomized to receive statins or placebo (e.g., as in CARDS \citep{colhoun2004primary}). 

In addition to a treatment effect, a decision-maker would also need an estimate of what would happen without a medication, $P(Y=1|do(T=0),x)$ (or even, although less useful, the marginal estimate, $P(Y=1|X)$).  If statins work (per their treatment effect), and the risk of an adverse outcome  without them is high, one would prescribe the medication.  By doing so, one would have heuristically addressed (\ref{eq:treatmentopt}). 
\end{theorem}

\begin{theorem}
\label{heur2}
One can also choose to avoid directly referencing a trial's treatment effect by using guidelines. For example, for statins, one can invoke the United States Preventive Task Force (USPTF) guidelines \citep{chou2022statin}, which recommend treatment when atherosclerotic cardiovascular disease (ASCVD) \citep{goff20142013} risk, $P(Y=1|x),$ is greater than some cutoff, such as $10\%.$
\end{theorem}

Note that the Heuristics \ref{heur1} and \ref{heur2} might not solve the true treatment problem, (\ref{eq:treatmentopt}), if, for example, there are other important outcomes, such as side effects, besides $Y$ (in such a case, it might be important to consider an approach, such as the one in \citet{thurston2009bayesian}, for multiple outcomes). Also, the heuristics described above might not solve the treatment problem, if the treatment does not work well for all patients. For example, this might occur when the conditional treatment effect, 
\begin{align}
\label{eq:cte}
\tau_x=E[Y|do(T=1),X=x]-E[Y|do(T=0),X=x],
\end{align} 
which---better aligned with the conditional nature of $P(u|do(t),x)$---varies as a function of $x$. For the statin example, this may occur if the efficacy of statins varies by age, and, therefore, a global cutoff of ASCVD $10\%$ may not be utility-maximizing in certain age groups. Also, Heuristic \ref{heur2} might not solve the treatment problem, because the ASCVD calculator, which was derived with observational data and does not adjust for all confounders, estimates $P(y|x),$ not $P(y|do(T=1),x).$ Finally, even if $P(y|do(T=1),x)$ is correctly estimated, Heuristic \ref{heur2}  might not solve the treatment problem, because the global ASCVD threshold may not reflect individual patient preferences.   

\subsubsection{Observational data}
\label{subsec:obsData}

In the same way that one can obtain an estimate for a treatment effect, $\tau,$  from a trial, one can, under certain conditions,  obtain an estimate of a treatment effect from an observational study \citep{pearl2009causality, rubin2005causal}. In particular, under causal assumptions---including the assumption that some set of variables $Z$ contains all confounders, which is known as the ``no unmeasured confounders assumption''---we have that \[E[Y|do(T=t)]=E[E[Y|t,Z]],\] where the latter can be estimated from  observational data.  For example, in the case of statins, one can theoretically estimate a counterfactual \citep{pearl2009causality} treatment effect from observational data (e.g., from the electronic health record), if one adjusts for all factors that influence both the patient's decision to take a statin and the cardiovascular outcome. These factors might include information such as age, comorbid conditions, etc. 

Note, however, that the no unmeasured confounders assumption cannot be verified with data alone. In other words, if we were to attempt to estimate the treatment effect of statins using observational data, we might be missing the confounder ``socioeconomic status,'' and, therefore, the estimated treatment effect would be incorrect, but we would not be able to determine this incorrectness using the data alone.

Overall, however, observational data gives us a way to---in theory---work around the need for randomization. Doing so is sometimes necessary for ethical reasons (e.g., one cannot randomize something like tobacco use). 
Note also that the conditional treatment effect described in (\ref{eq:cte}) has also been considered in the observational data setting (see, e.g., \citep{kennedy2023towards}).   It is also thought that observational data might complement trials for a variety of other reasons \citep{black1996we, van2025revolutionizing, concato2004beyond, ashley2024randomized}.  

In any case, observational data analysis requires strong, difficult-to-verify assumptions.

\subsubsection{Using observational data to directly solve the treatment problem}

Analogously to estimating $P(u|do(t),z)$ in a randomized controlled trial and then directly solving the treatment problem, (\ref{eq:treatmentopt})---as opposed to using Heuristics \ref{heur1} or \ref{heur2}---one can also directly solve (\ref{eq:treatmentopt}) using observational data---as opposed to heuristically solving it with the observational data treatment effect.   

Solving the treatment problem in this way involves
\begin{align}
\label{eq:measureChange}
\mathbb{E}_{\pi}U
\notag &=\int\int\int  u dP(u|do(t),z)d\Pi(t|z)dP(z)\\
\notag &=\int\int\int  u dP(u|do(t),z)d\Pi(t|z)\frac{d\Pi_0(t|z)}{d\Pi_0(t|z)}dP(z)\\
&=E_{\pi_0}\left\{
\frac{\pi}{\pi_0}U\right\},
\end{align} 
where $\Pi_0$ is the CDF of $\pi_0,$ the policy that generated the observed data (i.e., $T\sim \pi_0(T|Z)),$ and $Z$ contains all confounders, so $dP(u|do(t),z)=dP(u|t,z).$

Under the necessary assumptions, the expression in (\ref{eq:measureChange}) can then be estimated counterfactually from observed data, as long as an estimator for the propensity, $\pi_0,$ is plugged in.  This framework encompasses trials, which are equivalent to (\ref{eq:measureChange}) with a treatment policy (propensity score), $\pi_0=0.5.$   

Maximizing the objective (\ref{eq:measureChange}) with respect to $\pi$ is known as  policy search \citep{ng2013pegasus}, offline-reinforcement learning \citep{relsparSIM,thomas2015safe}, or importance sampling  \citep{precup2000eligibility, mcbook}.  It is closely related to inverse probability weighting \citep{horvitz1952generalization, robins1994estimation}.
Some variation of (\ref{eq:measureChange}) has been studied for medical problems  in the causal inference \citep{murphy2003optimal, 
luckett2019estimating, Chakraborty2013, yazzourh2025medical, laber2014dynamic} and the off-policy reinforcement learning \citep{futoma2020popcorn, raghu2017deep, raghu2017continuous, gottesman2020interpretable, sutton2018reinforcement, jonsson2019deep, petersen2019deep} literature. 

 It is difficult to say whether solving the policy search objective in (\ref{eq:treatmentopt})  would lead to full automation of treatment decisions.  It depends on observational data, and, as discussed above, there are strong assumptions associated with this, some of which are not verifiable with the data alone.  

Further, (\ref{eq:treatmentopt}) requires correct utility specification, as well as assumptions about homogeneity of utilities across patients, which may not hold for many problems.  For example, patients may differ in terms of their willingness to tolerate statin-induced side effects. Trials may suffer from utility heterogeneity as well (for example, this will be the case if a trial only assesses the rate of statin-induced side effects without a corresponding patient utility).  

Nevertheless, the objective (\ref{eq:treatmentopt}), in a way that is analogous to obtaining a treatment effect from observational data, might one day allow us to avoid randomization while still directly solving the treatment problem.  E.g., for the statin example, as long as all assumptions are met, $\pi^*=\arg\max$(\ref{eq:measureChange}) could be estimated purely from observational data (e.g., outpatient electronic health record), and would give a mapping from patient covariates, $X,$  such as age, cholesterol, etc,  described in (\ref{eq:X.statins}), to a statin treatment decision, $T.$

\subsubsection{Challenges of evidence-based medicine}
\label{sec:ChallengesEBM}

Overall, although evidence-based medicine takes us in the right direction, it is not a panacea. For example, beyond the issues mentioned in \ref{subsec:obsData}, trials can suffer from other things, such as small sample sizes and a lack of generalizability.  Observational studies cannot easily fix this, because, as mentioned, they require strong assumptions. Sometimes, we  even encounter observational data within trials, as in the case of non-adherence \citep{frangakis2002principal}.  Many studies also face difficulties related to choice of statistical analysis, significance, multiplicity, etc \citep{harrell2019bayesian, teira2011frequentist}. 
Further, even if studies are well-designed, applying their insights to clinical scenarios can be a challenge \citep{
keegan2023diversity, 
feinstein1997problems, deaton2018understanding, 
weisenthal2025information, 
rothwell2005external, 
ratnani2023evidence, 
feller2020judgment, donnellan2013health, henry2006recognizing, upshur2002if, herbert2001evidence, fava2006intellectual, tonelli2012official, braude2009clinical, 
huyuk2024adaptive, 
chin2018clinical}. 

Despite these potential issues, evidence-based medicine is the gold standard for addressing the treatment problem. 

\subsection{Experience}
\label{sect:experience}
For many medical decision problems, however, evidence-based studies might not exist. For example, we might not have good data on how statins work for young, otherwise-healthy patients with hyperlipidemia. 

In such a case, one must use other means to solve the treatment problem, (\ref{eq:treatmentopt}). For example, one could use clinical intuition \citep{greenhalgh2002intuition}
to determine if a patient will have a cardiovascular event; i.e., one would intuit $P(Y=1|do(T=0),x)$ or $P(U=u|do(T=0),x)$ based on patients seen in the past.  For example, a clinician might have seen that patients who do not take statins tend to have more adverse events, such as strokes, and this might give intuition that $P(\text{stroke}|\text{do(no statin)},\text{age},\cdots)$ is large. 

Note that estimating  $P(u|do(t),x)$ via experience requires an ability to tabulate events into degrees of belief, or frequencies. This might involve creating and populating a mental database  (such structure would ostensibly need to be programmed into a chatbot, if its engineers would like it to learn from experience), either via trial-and-error or observation.  The former is  precluded for the most part by ethics (e.g.,  one cannot (except within a trial), prescribe a statin for the sole purpose of seeing if it works), so, at the level of the individual clinician, observation is usually used (e.g., after having decided whether to prescribe a statin based on patient information and preferences, one observes the outcome and uses this to inform one's decisions with respect to future patients). As discussed in Section \ref{subsec:obsData}, however, observation is subject to confounding (which would occur, e.g., if patients who forego statins are usually sicker than those who take them). 
It is interesting that human clinicians can rely so much on observation in their own practice, despite the strong assumptions needed to formally analyze observational data, and still generally make reasonable treatment decisions---although this may be because observation is,  as we will now discuss, often augmented by mechanistic knowledge (and possibly also other components of causality \citep{hill1965environment}, such as dose-response, etc.).

\subsection{Knowledge}
\label{sec:MedKnowledge}
In practice, any treatment strategy depends on some mechanistic understanding of disease.\footnote{For an interesting discussion on knowledge in the context of clinical decision making, see \citet{tonelli2006integrating}.} For example, this understanding might be that statins decrease cholesterol via enzyme inhibition and that cholesterol is associated with vascular obstructions that lead to cardiovascular events, and, therefore, that one should prescribe statins when a patient has high cholesterol. 

Fundamentally, this kind of mechanistic knowledge allows clinicians to make statements about the likelihood of disease progression (cholesterol increase the likelihood of blockages and cardiovascular events), or treatment effects (statins inhibit enzymes and thus decrease the likelihood of cholesterol), without relying on direct experience (or evidence, which is experience in the collective sense), randomized or observational.  

Note that the results of a trial, i.e., treatment effects such as $\tau,$ are also a form of knowledge, and, hence, the boundary between knowledge and evidence may be blurred.

\subsection{Imitation}
\label{sec:informalImitation}

We can also imagine a clinician who attempts to solve the treatment problem,  (\ref{eq:treatmentopt}), by just imitating others.  The imitated treatment decision, $T,$ would be
\[T \sim {\hat{\pi}}(T|X)\]
where 
\begin{align}
\label{eq:imit.tx}
\hat{\pi}=\arg\min_{\pi}KL (\pi, \pi_0),
\end{align}
where KL is Kullback-Leibler divergence \citep{kullback1951information} and $\pi_0$ is the policy that is being imitated.  

If (\ref{eq:imit.tx}) is estimated with data, solving it is known as imitation learning/behavioral cloning \citep{kober2010imitation, torabi2018behavioral}, and, in such a case, it would be imitation of $\pi_0,$ the policy that generated the dataset (i.e., the clinicians in the hospital that generated the dataset were acting according to $\pi_0$).  Imitation learning is also known as  maximum likelihood estimation, since minimizing KL divergence is equivalent to maximizing likelihood \citep{van2000asymptotic}. 
To employ such an imitation strategy, one might  recommend a statin for a patient with high cholesterol, because that is what other clinicians usually do. 

Although imitation might reflect evidence when those imitated used evidence,  imitation is often not ideal for solving (\ref{eq:treatmentopt}).  For example, if a medical community does not know how to treat a certain disease, imitation of clinicians within this medical community will be problematic. The imitation objective in (\ref{eq:imit.tx}) does not take into account utility, $U,$ which is the essence of (\ref{eq:treatmentopt}). For example, imagine a hypothetical medical community in which clinicians only prescribe statins based on cholesterol levels. Then imitating clinicians in this community, i.e., solving (\ref{eq:imit.tx}) alone, would not take into account the final cardiac outcomes (or even the final cholesterol levels). 

Imitation, therefore, cannot solve the treatment problem directly.  This said, in medicine, imitation---to some degree---is important, since diverging from standard practices can be problematic.
 
\subsection{Combining imitation and patient utility optimization}
\label{sec:treatTRustREgion}

One can interpolate between imitation and utility maximization using methods from \citep{schulman2015trust}.  In particular,
\begin{align}
\label{eq:treatopt.trustregion}
\tilde{\pi}^* = \arg\max_{\pi}\mathbb{E}_{\pi}U(T,X) - \lambda d(\pi,\hat{\pi}),
\end{align}
where $d$ is some norm. A variation of (\ref{eq:treatopt.trustregion}) is considered for treatment problems (e.g., in \citet{futoma2020popcorn} and \citet{relsparSIM}) and non-treatment problems (e.g., in \citet{farahmand2016value} and \citet{ueno2012weighted}). Using this strategy for the statin example, one can make a treatment choice by considering the expected utility of prescribing the medication alongside what other clinicians might do in similar situations.

It is difficult to know how to choose $\lambda,$ which informs the tradeoff between imitation and utility maximization.  However, in practice, a clinician routinely weighs their own treatment recommendations against prevailing practices, and, in the process, sets $\lambda$. In other words, if a particular patient wants to avoid side effects at all costs, the clinician might recommend against starting a statin, which, if most clinicians would prescribe a statin, implicitly requires fixing $\lambda=0.$

Overall, medical decision making is a difficult task, and it makes sense that evidence-based medicine has emerged to help. Decision aids that help in other ways would ostensibly be welcome.   

We will now discuss chatbots and then the potential for language models to help with solving the treatment problem, (\ref{eq:treatmentopt}).  

\section{The chat problem}

Although the details of specific chatbots are not always made public, systems that solve the chat problem are comprised of language models trained within reinforcement learning frameworks \citep{ouyang2022training}. The reinforcement learning component is structurally similar to (\ref{eq:treatopt.trustregion}), which we discussed in terms of the treatment problem in Section \ref{sec:treatTRustREgion}.  Therefore, there will be many analogies between treatment and chat. However, as we will show, an agent that solves the chat problem cannot necessarily solve the treatment problem.

\subsection{Imitating chat}
We started our discussion of the treatment problem with utility optimization, but we will start our discussion of the chat problem with imitation.  We do so because one can think of chatbot engineers as prioritizing the Turing test \citep{turing2009computing} (a test for ``artificial intelligence''), which is, at some level, a test for whether a machine can seem human-like.   

If $Q$ is a question, or prompt, and $A$ is an answer, a language model that imitates a human solves the following problem,
\begin{align}
\label{eq:policy.imit.chat}
A \sim {\hat{\pi}}_c(A|Q),
\end{align}
where 
\begin{align}
\label{eq:imit.chat}
\hat{\pi}_c=\arg\min_{\pi_c}KL (\pi_c, {\pi_0}_c),
\end{align} where ${\pi_0}_c$ is the mapping that generated the observed text data (e.g., a set of human-generated question-answer pairs from the internet or another textual source), and the subscript $c$ denotes ``chat'' to distinguish $\pi_c$ from the aforementioned treatment policy, $\pi,$ in (\ref{eq:treatmentopt}). 

 A system that solves the chat problem (\ref{eq:imit.chat}) might appear to have human-like qualities, and humans appear to solve the treatment problem, (\ref{eq:treatmentopt}), but (\ref{eq:imit.chat}) and  (\ref{eq:treatmentopt}) are quite different.  

For example, when given a medical scenario and the question of whether to begin a statin, a system that solves (\ref{eq:imit.chat}) will, in a sense, compare this to a representation of similar queries in its training data and give a response that is as similar as possible to those responses. In terms of the statin problem, if the training data mostly involved online discussions of statin side effects, the response from the system will likely emphasize those side effects. If the majority of the training data involved textbooks on human physiology, the system would likely emphasize the mechanism of action of statins. Finally, if the training data involved mostly medical guidelines, the response would likely involve these guidelines. While the latter might be reasonable in some cases, solving  (\ref{eq:treatmentopt}) requires a formal calculation, which is different from all the scenarios mentioned above.

\subsection{User utility}
\label{sec:userutil}
 We will now introduce the idea of maximizing user utility, or user preference, a problem which is analogous to \ref{eq:treatmentopt} and might be considered the primary \textit{commercial} chat problem.

In solving the user utility problem, an engineer designs a system that learns a distribution from which an answer, $A,$ rather than---as in (\ref{eq:treatmentopt})---a treatment, $T,$ is drawn, and this is done conditionally on a prompt, $Q,$ rather than on patient information, $X.$ In particular,  the response, $A,$ which maximizes user utility, is drawn as
\begin{align}
\label{eq:chatpolicy}
A \sim \pi^*_c(A|Q),
\end{align}
where the subscript $c,$ as above in (\ref{eq:imit.chat}), denotes chat (in order to distinguish this policy from the  treatment policy, $\pi,$ which was mentioned in  (\ref{eq:treatmentopt})).

The distribution $\pi^*_c$ is obtained as  
\begin{align}
\label{eq:chatopt}
\pi^*_c = \arg\max_{\pi_c}\mathbb{E}_{\pi_c}S(A,Q),
\end{align}
where $\mathbb{E}_{\pi_c}S$ represents, similarly to (\ref{eq:detail}), 
\begin{align}
\label{eq:detailChat}
\mathbb{E}_{\pi_c}S&=\int sdP_{\pi_c}(s)\\&=\int\int\int  s dP(s|do(a),q)d\Pi_c(a|q)dP(q),
\end{align} with $\Pi_c$ the CDF of $\pi_c,$ and $S$ denoting user utility with respect to the chatbot's answer. Then $\mathbb{E}_{\pi_c}S(A,Q)$ is the expected human user's utility under arbitrary policy, $\pi_c$.  

As an aside, one can also imagine a sort of post-answer state, $Q'$ (the post-answer state of the user), analogous to post-treatment state, $X',$ discussed in (\ref{eq:detail}). However, as with (\ref{eq:detail}), we suppress  $Q'$ from (\ref{eq:detailChat}), since $S(Q')$ is a function of it.

For the statin example, one can imagine that $Q$ contains a description of a patient (and the question of whether or not to start a statin), and $A$ contains an answer that the user finds most satisfactory (it could be, for example, the most human-like answer, the most agreeable answer, etc).

Note that (\ref{eq:chatopt}) and (\ref{eq:treatmentopt}) are similar in that they are both optimization problems.  However, they have very different inputs, outputs, and utilities.  Therefore, maximizing chatbot user utility does not imply the ability to maximize patient utility. 

If a system is developed to solve (\ref{eq:chatopt}), its medical recommendations might change when a user is asking from different perspectives (which has been confirmed empirically with respect to clinicians vs insurance agents \citep{yu2024medical}). In contrast, the treatment problem (\ref{eq:treatmentopt}) has one optimal solution (for each patient utility function), regardless of who is asking the question, because $dP(u|do(t),z)$ in (\ref{eq:detail}) is fixed. 

For the statin example, if the users whose preferences were used to train the system were predominantly from a more lifestyle-oriented community, the system would likely favor diet and exercise over statins. However, the  probabilities that lifestyle measures (e.g., diet, exercise) and statins will work are fixed, and one of the two (or some combination) is necessarily a better recommendation, depending on patient information and preferences (as opposed to the preferences of the users who were employed to train the system).


\subsection{Evidence}
\label{section:Chatevidence}

To truly maximize user utility, $S,$ a chatbot is programmed to experiment with respect to its response, $A$.  This is called ``exploration" in reinforcement learning, and it is analogous to ``randomization'' in evidence-based medicine.  In other words, chatbot engineers (not the chatbots themselves, which would align more with what is described in \citep{silver2025welcome}) are constantly running some version of randomized-controlled trials and then updating the chatbots with the results.   
It is important to remember that chatbot engineers have greater flexibility than clinical decision aid engineers in terms of experimentation in this way, because often the stakes in (non-healthcare-related) chat are (in theory, assuming the non-healthcare-related chat does not involve major life advice) lower than in healthcare.  

Chatbot engineers might also incorporate varying degrees of observational data, via off-policy reinforcement learning, analogous to the observational data treatment problem in (\ref{eq:measureChange}). In particular, in doing so, they estimate
\begin{align}
\label{eq:measureChangeChat}
\mathbb{E}_{\pi_c}S
=E_{\pi_{0c}}\left\{
\frac{\pi_c}{\pi_{0c}}S\right\},
\end{align} 
which depends on $\pi_{0c}$  instead of $\pi_0,$ as did (\ref{eq:measureChange}), where the subscript ``c'' is again used to designate ``chat."  Estimating (\ref{eq:measureChangeChat}) depends on observational data, but, as mentioned above, the chat problem is often lower stakes, so the violation of any observational data analysis assumptions may go unnoticed. 

Overall, collecting evidence for the chat problem, whether this evidence be from randomization or observation, is often easier than collecting evidence for the treatment problem, because the chat problem is usually lower stakes.

\subsection{Combining imitation and user utility optimization}
Chatbots are trained to limit their divergence from the original imitation policy, $\hat{\pi}_c,$ using methods similar to those in trust region policy optimization \citep{schulman2015trust}.  Hence, there is a tradeoff between imitation and utility-maximization that is similar to the tradeoff discussed in (\ref{eq:treatopt.trustregion}) for the treatment problem. In particular, we have 
\begin{align}
\label{eq:chatopt.trustregion}
\tilde{\pi}^*_c = \arg\max_{\pi_c}\mathbb{E}_{\pi_c}S(A,Q) - \lambda d(\pi_c,\hat{\pi}_c),
\end{align}
where $d$ is some norm.  By solving the tradeoff optimization in (\ref{eq:chatopt.trustregion}), a system like ChatGPT can incorporate user preferences while still appearing human in its responses (in contrast, a recommender system \citep{leblanc2024recommender} maximizes for user preference, but it is not conversational). 

As an aside, besides allowing for the tradeoff between utility and imitation, \citep{schulman2015trust} was developed to stabilize utility optimization (and it has other functions related to safety \citep{achiam2017constrained});  in practice, when training chatbots, for computational reasons, the Lagrangian form in (\ref{eq:chatopt.trustregion}) is likely replaced by its extension in \citet{schulman2017proximal}.   

Out of the box, a model that solves (\ref{eq:chatopt.trustregion}) will not, in a formal sense, solve the treatment problem (\ref{eq:treatmentopt}). The differences between $\pi^*$ and $\tilde{\pi}_c^*$ are significant, because $S$ is different from $U$, $A$ from $T$, and $Q$ from $X.$  For the statin example, $S$ would be the average user's satisfaction with the system's response, $Q,$ as it relates to statins, whereas $U$ would reflect the average patient's medical outcomes (cardiovascular outcomes, side effects), and the corresponding utility, as it relates to treatment choice $T.$ Also, $A$ would be the system's textual response, whereas $T$ would be a treatment choice with respect to statins. Finally, $Q$ would be the patient's input to the chatbot, while $X$ would be the patient's covariates.

Further, solutions to the chat problem appear to fundamentally prioritize imitation, as in (\ref{eq:imit.chat}), instead of user utility, as in (\ref{eq:chatopt}), where optimizing for the latter is often described as a sort of additional``fine-tuning'' step \citep{ziegler2019fine}.  For the statin example, a chatbot trained according to (\ref{eq:chatopt.trustregion}) would tend to imitate the statin-related (and other-topic-related) question-answer samples from the training data rather than optimizing for user preference.  Either way, neither imitation nor optimizing for user chat preference solves the treatment problem.

\subsection{Knowledge}
\label{sect:mindful}
Within a chat, a user might request that a chatbot's responses change in some way. For example,  the user might request that the chatbot “be more mindful,” in which case the chatbot can search for a book on mindfulness, read it (or access its own understanding of the concept), and change its style accordingly. Or, in terms of the statin example, a user might request that a chatbot take a more lifestyle-oriented approach when discussing hyperlipidemia.

This capability for within-chat behavioral modification can be useful for many problems in medicine (e.g., for adapting to a patient's conversational or content-related preferences).  As discussed in Section \ref{sect:experience}, however, a chatbot cannot use knowledge obtained in this way to permanently change its internal parameters, independent of its engineers---any changes will be erased when the chat ends (in contrast, a human can change their internal parameters based on new information). In other words, for the statin example, if a user asks a chatbot to be more lifestyle-oriented with respect to hyperlipidemia treatment, this does not mean the chatbot will be so for the next user.  

A chatbot does have an internal store of knowledge in its parameters. This knowledge is, by design, not necessarily factual (it might be better described as ``approximate''\citep{KareemHall})---parameters depend on the objective functions for which they are the solutions, and (\ref{eq:imit.chat}) and (\ref{eq:chatopt.trustregion}) do not optimize for factual accuracy.  However, it can be somewhat factual.  

\subsection{Other uses}

Chatbots might not need to directly solve (\ref{eq:treatmentopt}) to be helpful in medicine.

\subsubsection{Medical knowledge}
\label{sec:chatLitSearch}
One could use a chatbot trained on (\ref{eq:chatopt.trustregion}) to aid with searching and critically interacting with the medical literature \citep{goh2024large, alaa2024large, zwaan2024cognitive}.  In fact, in doing so, a chatbot trained to solve (\ref{eq:chatopt.trustregion}) may appear to act autonomously as a clinician---it might field a question, search the literature, and develop a guideline-based treatment plan.  For the statin example, given a patient's ASCVD score, the chatbot might access USPTF guidelines to tailor its statin recommendation. This is similar to how a chatbot might search the literature for instruction on how to emphasize lifestyle choices, as discussed in Section \ref{sect:mindful}.  Although human clinicians likely solve some medical problems by taking a knowledge-based, rather than directly experience-based or evidence-based approach, one should be cautious when using chatbots in this way, as shown in \citep{polzak2025can}.   

Ultimately, it is important to remember the difference between (\ref{eq:chatopt.trustregion}) and (\ref{eq:treatmentopt}); a chatbot, even if it searches the medical literature and develops a plan based on a guideline, is not solving the treatment problem itself. Using a chatbot trained for (\ref{eq:chatopt.trustregion})  to solve (\ref{eq:treatmentopt}) is indirect. Doing so is similar to a human investigator who would like to know the average blood pressure of a room of people asking a chatbot to guess it. The chatbot, with the internet at its disposal, might give a good guess. However, the investigator  should have instead recorded each person's blood pressure and taken an average, which is analogous to solving the treatment problem in (\ref{eq:treatmentopt}) directly (via, e.g., a well-designed trial or using (\ref{eq:measureChange}), if possible).   In this case, a calculator (to calculate the average) would  be superhuman (perfect) in its estimate of the average, outperforming a chatbot's guess (although a chatbot's guess might outperform a human's guess). 

Overall, also, the treatment problem, (\ref{eq:treatmentopt}), is more complex than taking an average.  This may be why human clinicians often solve it heuristically, and, interestingly, why a chatbot that solves it heuristically might sometimes seem to hold its own (as in the just-mentioned blood pressure example, it may be better than a human). Further, a system or clinician that attempts to solve  (\ref{eq:treatmentopt}) directly might make unrealistic assumptions.  In other words, if a clinician does a true risk-benefit decision analysis for the statin problem, but their estimate of the probability of side effects is incorrect, they may underperform in terms of decision-making relative to a clinician who does not use a decision-analysis framework. However, in terms of the statin example, statins are fairly well studied, and not  prescribing them in an evidence-based way would likely be, on average, a subpar strategy. One approach might be to use chatbots within a decision-theoretic framework; for example,  chatbots could be used to (verified for accuracy) efficiently list the risks and benefits of statins, which could be used by the clinician and patient to arrive at a decision.

\subsubsection{Support}
Many problems---especially those related to prevention and chronic disease---benefit from gestalt knowledge, availability, and patience.  For example, narrative medicine is a well-established approach (see e.g. \citep{egnew2009suffering}), and  a chatbot might---via qualities like patience and availability---support a patient to develop a narrative.  For the statin example, a patient might be reluctant to start a statin due to past difficulties with other medications, or they may want to solidify their own preferences with respect to the possible outcomes associated with statin treatment, such as side effects; a clinician might not (based on scheduling protocols beyond their control) be able to fully engage in these conversations, whereas a chatbot might.  Note that a chatbot's conversational qualities in this way, however, will always be a function of its reward, which is specified by its engineers. 

It is difficult to say at what point a medical problem becomes a good candidate for a chatbot's involvement---perhaps low-stakes, transparent problems, which allow for experimentation and facilitate user awareness of potential errors, respectively, would be suitable.  The statin treatment problem does not meet these criteria, but other problems might. Either way, patients will likely begin to use chatbots for medical purposes long before any official regulatory approval.

Generally, though, asking a model trained according to (\ref{eq:chatopt.trustregion}) to support a patient, or to help with searching or interpreting the literature---although potentially beneficial---is not the same as directly solving the treatment problem, (\ref{eq:treatmentopt}).

\section{Language models and the treatment problem}
Below, we will discuss how an engineer might train a chatbot-like system specifically to solve the treatment problem, (\ref{eq:treatmentopt}). However, first, we will briefly discuss the imitation of medical notes.

\subsection{Medical note imitation}
One might decide to  train a language model with a large number of medical notes (e.g., from the electronic health record, assuming high data volume and quality). The language model would take as input patient information,  $I$ (e.g., the subjective, objective, and assessment sections of a note---which, for the statin example, might indicate that the patient is presenting with a family history of cardiac events, tobacco use, and elevated cholesterol)---and provide as output a treatment plan, $\tilde{P}$ (learned according to the plan section of the electronic health record notes--- which, for the statin example, might include a recommendation that the patient take the medication).  In other words, one would draw a treatment plan, $\tilde{P},$
\begin{align}
\label{eq:policy.imit.text.treat}
\tilde{P} \sim {\hat{\pi}_{tc}}(\tilde{P}|I)
\end{align}
where
\begin{align}
\label{eq:imit.chat.tx}
\hat{\pi}_{tc}=\arg\min_{\pi_{tc}}KL (\pi_{tc}, \pi_{0tc}),
\end{align} where ${\pi_0}_{tc}$ is the data-generating mapping from patient information to a plan (i.e., the human clinicians who wrote the original notes were following $\pi_{0tc}$), and the subscript $tc$ stands for ``treatment, chat'' to distinguish $\pi_{0tc}$ from $\pi_{0c}$ in the imitation chat problem, (\ref{eq:imit.chat}).  For example, the policy $\pi_{0tc}$ could be estimated using a dataset that contains notes on patients for whom statin decisions were made.

Solving  (\ref{eq:imit.chat.tx}) is equivalent to the treatment imitation problem in (\ref{eq:imit.tx}) (\textit{not} the  chat imitation problem in (\ref{eq:imit.chat})), but it uses text data, $I,$ instead of $X,$ and $\tilde{P},$ instead of $T.$ In other words, the text data of the note itself serves as the input rather than tabular data indicating smoking=1, cholesterol=233, etc. One could also augment $I$ with laboratory data, images, etc.  For the statin example, one could include the patient's cholesterol levels, stress test results, etc.

However, either way, as mentioned in Section \ref{sec:informalImitation}, imitation has its own drawbacks. In particular, if the standard of care, $\pi_{0tc},$ is suboptimal, then its imitation will be the same (e.g., if the USPTF guideline for statin administration in Heuristic \ref{heur2} does not always solve the treatment problem in an optimal way, then imitating clinicians who follow it will not lead to an optimal solution, either).  As mentioned in Section \ref{sec:informalImitation}, imitation lacks a utility signal,  which is the essence of the true treatment problem, (\ref{eq:treatmentopt}).  In terms of the statin problem, the pure imitation treatment strategy described above integrates neither patient outcomes (e.g., cardiovascular events or statin side effects) nor subjective patient experiences of these outcomes, which are at the heart of the statin treatment problem.

\subsubsection{Disconnect between appearance and substance}
\label{sec:TermDiscHype}
A system that solves   (\ref{eq:imit.chat.tx})  will likely have sophisticated command of medical vocabulary, and it  may appear to solve the treatment problem (in an evidence-based way, if the notes discuss evidence).  This is a facade.  %
This difference between appearance and substance, perhaps even more so than the one described in (\ref{sec:chatLitSearch}), can be dangerous in a field such as medicine. 

In general, such disconnects between appearance and substance appear frequently with respect to language models (e.g., see \citep{hao2025medpair}).  Possibly such disconnects are due to the terms, such as ``artificial intelligence,'' used to describe chatbots, which often involve anthropomorphism of mathematical concepts. Although anthropomorphism generates interest, it also obscures connections between chatbots and established approaches (e.g.,  consider how we have shown the connection between (\ref{eq:treatmentopt}), which is an established problem, and (\ref{eq:chatopt}), which is often called ``artificial intelligence,'' by writing the equations rather than relying on terminology alone).  Losing these connections in a high-stakes field like medicine can be problematic. For example, one might argue that the public has a healthy skepticism for statistical  modeling; (for the statin example, patients are rightly skeptical when one performs an ASCVD calculation and then states that, based on the results, a statin is necessary), but patients are not as skeptical about the more vaguely-defined ``artificial intelligence," even though the essences of ASCVD and a language model, in particular as related to potential drawbacks (bias, variance, extrapolation, etc), are very similar.  Healthy skepticism is essential for patients, clinicians, and funding agencies, because it helps them understand how these new technologies might impact care. 

\subsection{Combining medical note imitation and patient utility}
\label{sec:LLMTx}
Ultimately, a system that truly solves the treatment problem, (\ref{eq:treatmentopt}), not just the imitation treatment problem (\ref{eq:imit.tx}),  requires a patient utility signal. In other words, for the statin example, training such a system would require that one take into account the final cardiovascular outcomes (and how the patient experiences them).

This signal could come from, for example, follow-up notes (for the statin example, engineers could look ahead 10 years after statin administration and see whether the patient had a cardiovascular event, as well as any side effects, and also how the patient ended up experiencing these outcomes from a utility standpoint).  Obtaining utility in this way might have issues, such as loss to follow-up, measurement difficulties, etc. However, these problems are fundamental to medical decision-making itself, language models or not (i.e., any decision aid will have to grapple with outcomes, utilities, loss to follow up). Language models, which have been used, for example, to extract complex signals from medical text \citep{miao2025understanding} might be useful in extracting utility from medical notes. In such a case, these notes might be considered the post-treatment state, $I',$ of which utility, $U(I'),$ would be a function.  The engineers would then solve for
\begin{align}
\label{eq:treatopt.trustregion.chat}
\pi^*_{tc} = \arg\max_{\pi_c}\mathbb{E}_{\pi_c}U(\tilde{P},I) - \lambda d(\pi_c,\hat{\pi}_{tc}),
\end{align}
where recall that $I$ is an input that contains information about the patient  (possibly a hybrid of notes and structured data), $\tilde{P}$ is a text-based (or numeric) output that reflects a treatment plan, and $d$ is a norm.  

This policy $\pi^*_{tc},$ could  fundamentally  focus on solving the treatment optimization problem in (\ref{eq:treatmentopt}) but also have an imitation component that prevents it from deviating too far from the clinicians who generated the medical notes (i.e. those who were acting according to $\pi_{0tc}$). One could imagine adding another penalty, $\lambda_c d(\pi,\hat{\pi}_{c}),$ to ensure that the system remains conversational, although $\lambda_c$ might be difficult to set.  

For the statin example, such a system would attempt to make a treatment decision that increases the patient's utility (which may differ according to individual preferences related to cardiovascular events and side effects), while also taking into account what is commonly done.

Either way, if we could solve (\ref{eq:treatopt.trustregion.chat}), we may be able to achieve results for the treatment problem (\ref{eq:treatmentopt}) that are similar to ChatGPT's (in many ways, superhuman) results on the chat problem (\ref{eq:chatopt}). In other words, for the statin example, in the same way one can use a calculator to compute the average blood pressure in a room full of people, and this computation can be expected to be optimal, the system would be able to compute the optimal decision with respect to statin administration. 

However, there is one major hurdle.  How to actually solve  (\ref{eq:policy.imit.text.treat})?   This brings us back to the discussion in Section \ref{sec:TreatevidenceExperience} on the treatment problem, (\ref{eq:treatmentopt}), as it relates to evidence-based medicine. In particular, we must again discuss trials (experimentation) and observation. 

\subsubsection{Experimentation}
Chatbots conceivably excel at maximizing user utility in the chat problem, (\ref{eq:chatopt}),  because they can be programmed to experiment with their responses \footnote{This experimentation is performed on paid beta-testers and also users. This likely does not bother the users too much, since they continue to use chatbots, although it is possible that the paid versions of chatbots have less experimentation than the unpaid versions.} and their engineers can then retrain the models according to the experiments. However, to solve (\ref{eq:treatopt.trustregion.chat}) in this way, a system would need to experiment on patients.  For example, to obtain the data necessary to train a system to solve the statin treatment problem, engineers would have to randomly assign statins to some patients and not to others, and then follow up on the outcomes. Although this is more difficult than following up on user chat experiences, the primary barrier is not logistical but ethical (this is the case for human clinicians too, and hence we have trials, and trials have strict oversight).  

This inability to experiment will hinder progress in terms of using reinforcement learning to train a system to solve the treatment problem, (\ref{eq:treatmentopt}), because reinforcement learning thrives in settings where experimentation is easy and can be done at scale \citep{silver2016mastering} (e.g., in chess, because all the rules are known, it is easy to simulate a million opponents and train the system by letting it play against them). 

For the statin problem, if one could accurately simulate patient physiology, such that virtual patients could be constructed and then statins assigned to them, and then the outcomes analyzed, it would be possible to develop a system that is able to solve the statin problem in a superhuman way, as systems can play chess or Go in superhuman ways.  The issue is that chess and Go operate according to fixed, programmable rules, whereas cardiovascular outcomes depend not just on highly stochastic physiology, but also randomness in other respects: for the statin example, a patient may randomly decide to start using tobacco, or to quit.  It is difficult to simulate this, and, therefore, experimentation on real subjects (in the form of clinical trials) is necessary.

Overall, if the success of systems like ChatGPT depends mostly on experimentation (and it is likely that this is the case), we are unlikely to be able to use the techniques used to train ChatGPT to create a system that treats disease as well as ChatGPT chats.  This is an ethical, not a technological, barrier. 

\subsubsection{Observation}

Interestingly, though, even though chat is not really a game like Chess or Go, ChatGPT still excels at chat.  This may imply that off-policy learning (on observational data), which was discussed in Section \ref{subsec:obsData}, contributes to its success.  I.e., ChatGPT is solving (\ref{eq:measureChangeChat}).  Using observational data in this way, one can, in theory, work around the need for randomization. For example, if such an approach is possible, one may be able to use observational, electronic health record notes and outcomes to develop a system that prescribes statins with superhuman optimality.

However, there are difficulties with observational data, such as positivity violations (more likely due to the high-dimensional nature of propensity $\hat{\pi}_{tc}$) and difficulty satisfying the no-unmeasured-confounders assumption.  For the statin problem, positivity violations might occur because no two patient scenarios are exactly alike, or because clinicians who generate observational data often use guidelines, which leads to limited variation in the observed treatment choices.  Unmeasured confounding may occur because many statin decisions depend on factors that are not recorded in the medical record (or are difficult to record accurately). 
 
 It is difficult to believe that assumptions such as positivity and no unmeasured confounders are satisfied for ChatGPT, but, as mentioned in Section \ref{section:Chatevidence},  violations, and their downstream effects, may be hidden behind the fact that a chatbot's errors\footnote{Note that these errors are likely different from what are called ``hallucinations,'' (see, e.g., \citep{alkaissi2023artificial}) although they may be related.} are difficult to detect and often harmless.  Unlike with chat, if a system attempting to solve the treatment problem, (\ref{eq:treatopt.trustregion.chat}), violates these assumptions, and therefore gives poor treatment decisions, this will be detected by patients, because it will cause harm. 

 With or without language models, determining the degree to which the treatment problem can be solved using observational data is an active problem in medical decision making (e.g., see the references in Section \ref{subsec:obsData}).   Interestingly, as a legal document, a medical note should, in theory, contain a complete justification for the management plan and, therefore, some representation of all confounders. However, the comprehensiveness of the justification varies.  

\subsection{Next steps for solving the treatment problem}

In discussing the treatment problem in Section \ref{sec:LLMTx},  note that we have found ourselves again on the topics of randomization and observation, which were  discussed in Section \ref{sec:TreatevidenceExperience} with respect to evidence-based medicine. Language models might help with the observational approach, as a sub-direction of the dynamic treatment regimes or off-policy reinforcement learning research areas (i.e., $\pi^*_{tc}$ in (\ref{eq:treatopt.trustregion.chat}) would be a language model), since the literature discussed in Sections \ref{subsec:obsData} and \ref{sec:treatTRustREgion} have traditionally considered structured instead of unstructured (text) data. It is worth exploring how language models, which are new, and text data, which is not, would impact this research area. However, this is a moonshot.  
It is vital that this high-risk, high-reward endeavor not overshadow (based on hype, which, as discussed in Section \ref{sec:TermDiscHype}, is likely due in part to anthropomorphic terminology) other important problems in evidence-based medicine, such as those discussed in Section \ref{sec:ChallengesEBM}, many of which are quite pressing.  

Interestingly, many of the problems with evidence-based medicine relate closely---via dependence on probability, optimization, etc---to chatbots.  Further, the new chatbot-invigorated interest in reinforcement learning (which is fundamentally linked to decision analysis) might help focus evidence-based medicine research on the patient's expected utility (often, studies have prioritized outcomes rather than utilities). Also, let us not discount the potential for chatbots to help in other ways (besides solving the treatment problem directly), such as by offering support to the patient or by acting as a co-pilot for the clinician when searching the literature. In order for this to happen, though, humans must still generate the literature, and, in particular, they must do so by conducting trials.

\section{Summary and conclusion}

Overall, medical decision making must start with maximizing patient utility. This is a difficult problem. In practice, it often requires that a clinician balance evidence, knowledge, experience, and imitation (e.g., a clinician who makes a healthcare decision about a statin uses risk scores,  literature on treatment effects, and their own clinical experience to assess risks and benefits).

It is important to understand that the chat problem, which focuses mostly on the imitation of conversation, is different from the treatment problem, which focuses mostly on the risks and benefits involved in optimizing patient utility. A chatbot,  by nature, does not solve the treatment problem. In a human-like way, it may tell a user whether or not to start a statin, but it does not perform the accompanying risk benefit calculations. This is not to say that a chatbot cannot promote health in other ways (i.e. by  helping clinicians search the literature or by supporting patients). However, these are not directly related to solving the true treatment problem.

Designing a system to solve the true treatment problem is complex  (imitating medical notes will not work).  The barriers  are less about technology and more about the ethics of experimentation.  One may be able to work around this using observational data, which is a known research area within evidence-based medicine. Language models---and text mining more generally---might help, but this is a moonshot problem. While language models might have potential here, it is important that the hype associated with ``artificial intelligence'' not overshadow other important problems in evidence-based medicine, many of which are quite pressing.  

\bibliographystyle{elsarticle-harv} 
\bibliography{bib}
\end{document}